\documentclass[aps,pra,showpacs,amsmath,amssymb]{revtex4-1}
\usepackage{graphicx}

\begin{document}

\title{Photonic band-gap properties for two-component slow light}

\author{J. Ruseckas}
\email{julius.ruseckas@tfai.vu.lt}
\homepage{http://www.itpa.lt/~ruseckas}

\author{V. Kudria\v{s}ov}

\author{G. Juzeli\=unas}
\email{gediminas.juzeliunas@tfai.vu.lt}
\homepage{http://www.itpa.lt/~gj}

\affiliation{Institute of Theoretical Physics and Astronomy, Vilnius University,
A.~Go\v{s}tauto 12, Vilnius 01108, Lithuania}

\author{R. G. Unanyan}

\author{J. Otterbach}

\author{M. Fleischhauer}

\affiliation{Fachbereich Physik and Research Center OPTIMAS,
Technische Universit\"at Kaiserslautern, 67663 Kaiserslautern, Germany}


\date{\today}

\begin{abstract}
We consider two-component ''spinor'' slow light in an ensemble of atoms
coherently driven by two pairs of counterpropagating control laser
fields in a double tripod-type linkage scheme. We derive an equation of
motion for the spinor slow light (SSL) representing an effective Dirac
equation for a massive particle with the mass determined by the
two-photon detuning. By changing the detuning the atomic medium acts as
a photonic crystal with a controllable band gap. If the frequency of the
incident probe light lies within the band gap, the light tunnels through
the sample. For frequencies  outside the band gap, the transmission
probability oscillates with increasing length of the sample. In both
cases the reflection takes place into the complementary mode of the
probe field. We investigate the influence of the finite excited state
lifetime on the transmission and reflection coefficients of the probe
light. We discuss possible experimental implementations of the SSL
using alkali atoms such as Rubidium or Sodium.
\end{abstract}

\maketitle

\section{Introduction}

\begin{figure}
\includegraphics[width=0.3\textwidth]{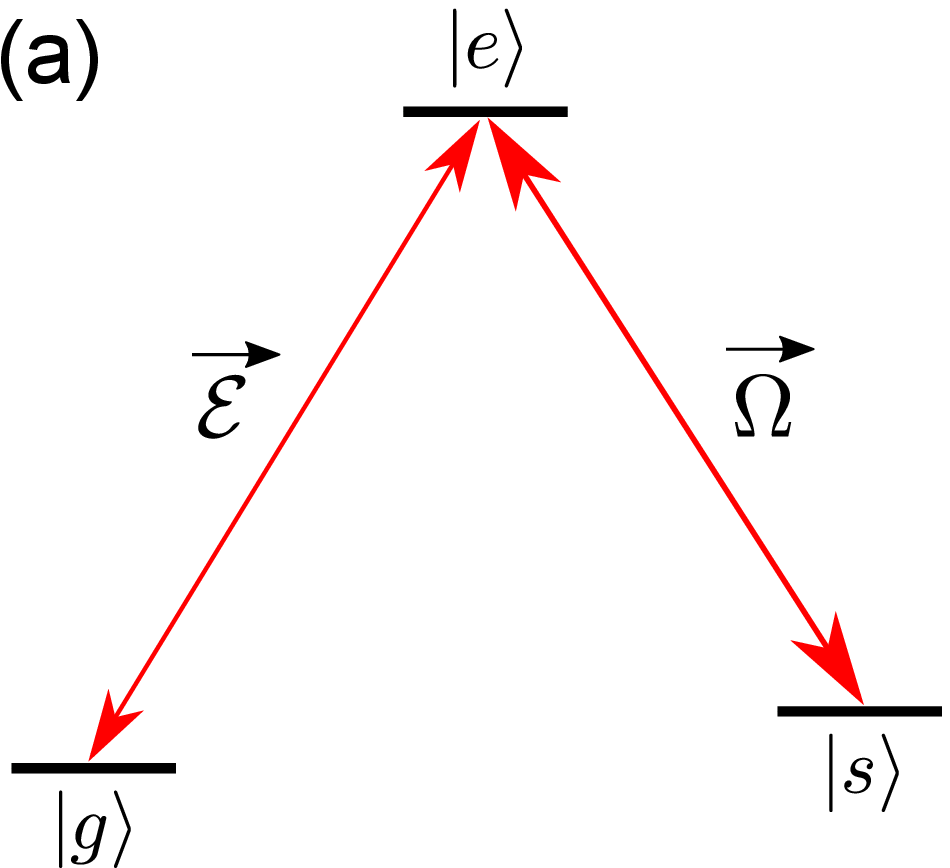}\includegraphics[width=0.3\textwidth]{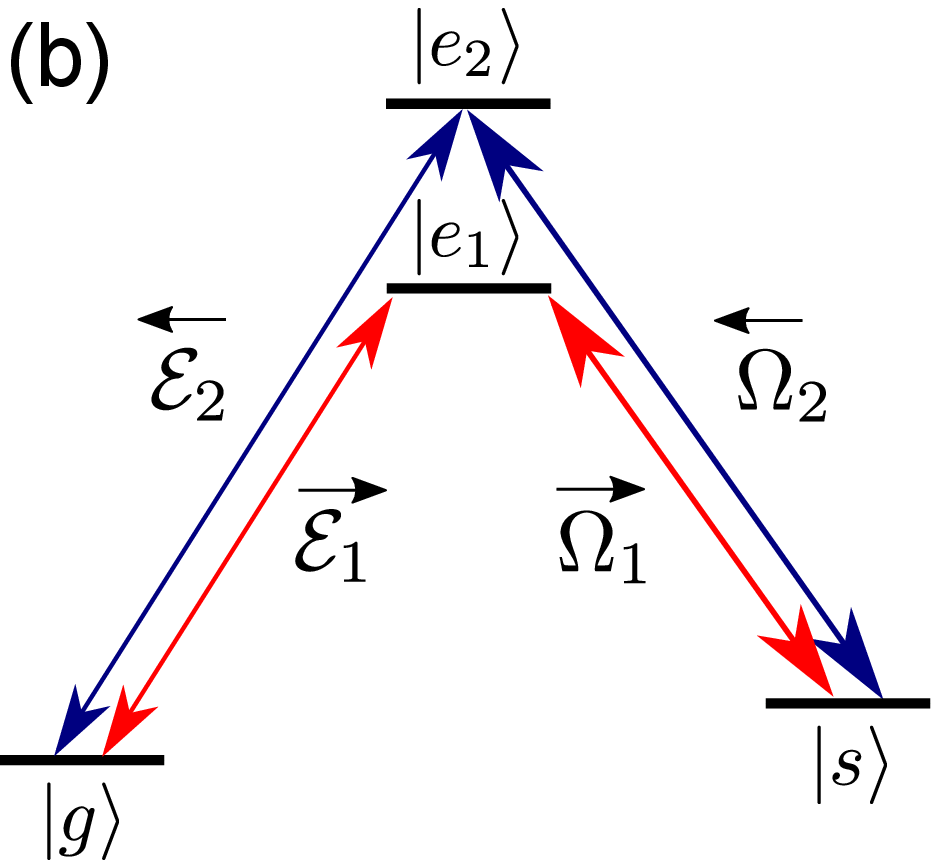}
\caption{(Color online) (a) The $\Lambda$ scheme of the atom-light
coupling involving a weak probe field $\mathcal{E}$ and a stronger
control field $\Omega$. Application of the control laser beam enables a
lossless propagation of the probe beam due to Electromagnetically
Induced Transparency (EIT); (b) The double $\Lambda$ setup for the
creation of single-component stationary light by using two
counter-propagating control fields $\Omega_1$ and $\Omega_2$ driving
different atomic transitions $|s\rangle\rightarrow|e_1\rangle$ and
$|s\rangle\rightarrow|e_2\rangle$, respectively.}
\label{fig:fig1}
\end{figure}

Over the last decade there has been a great deal of interest in slow
\cite{Hau-Nature-1999}, stored
\cite{Phillips-PRL-2001,Liu-Nature-2001,Ginsberg-Nature-2007,Schnorrberger-PRL-2009,Zhang-PRL-2009,Firstenberg-NaturePhys-2009}
and stationary \cite{Bajcsy-Nature-2003,Lin-PRL-2009} light. Coherent
control of slow light leads to a number of applications, such as
generation of non-classical states in atomic ensembles and reversible
quantum memories for slow light
\cite{Fleischhauer-PRL-2000,Juzeliunas-PRA-2002,Zibrov-PRL-2002,Lukin-RMP-2003,Eisaman-Nature-2005,Appel-PRL-2008,Honda-PRL-2008,Akiba-NJP-2009},
as well as non-linear optics at low intensities
\cite{Schmidt-OL-1996,Harris-PRL-1999,Fleischhauer-RMP-2005}.
Furthermore, propagation of light through moving media
\cite{Leonhardt-PRL-2000,Ohberg-PRA-2002,Fleischhauer-PRL-2002,Juzeliunas-PRA-2003,Artoni-PRA-2003,Zimmer-PRL-2004,Zimmer-2006a,Padgett-OL-2006,Ruseckas-PRA-2007}
can be used for rotational sensing devices. Slow light is formed in an
atomic medium with a $\Lambda$-type linkage pattern
(Fig.~\ref{fig:fig1}a) under conditions of Electromagnetically Induced
Transparency (EIT)
\cite{Arimondo-PO-1996,Harris-PT-1997,Scully-Book-1997,Lukin-RMP-2003,Fleischhauer-RMP-2005}.
The $\Lambda$-scheme involves two atomic ground states and an excited
state, as shown in Fig.~\ref{fig:fig1}a. EIT emerges due to the
destructive interference between atomic transitions from different
ground states to a common excited state induced by a weak probe beam and
a stronger control beam
\cite{Arimondo-PO-1996,Harris-PT-1997,Lukin-RMP-2003,Fleischhauer-RMP-2005}.
EIT allows to transmit a resonant probe beam through an otherwise opaque
atomic medium coherently driven by a control laser field and forms the
basis of many interesting applications as e.g.\ creating stationary
excitations of light
\cite{Moiseev-PRA-2006,Zimmer-OC-2006,Zimmer-PRA-2008,Fleischhauer-PRL-2008,Otterbach-PRL-2010}
in more complex double $\Lambda$ schemes as shown in
Fig.~\ref{fig:fig1}b, Bose-Einstein condensation of photons
\cite{Fleischhauer-PRL-2008,Zimmer-arXiv-2011}, or artificial magnetic
fields \cite{Marzlin-PRA-2008,Otterbach-PRL-2010} for photons.

It is to be pointed out that both ordinary and double $\Lambda$ schemes
support a single component slow and stationary light driving a single
atomic coherence $|g\rangle\rightarrow|s\rangle$. By adding an
additional control laser which couples an excited state to an additional
ground state, one arrives at a tripod linkage pattern
\cite{Unanyan-OC-1998} characterized by two atomic coherences. However,
in that case the slow light excitations remain in a single component,
because the original and additional control laser beams induce
transitions to a special superposition of the atomic ground states and
thus effectively drive a single atomic coherence
\cite{Paspalakis-PRA-2002,Raczynski-OC-2006,Raczynski-PRA-2007,Ruseckas-PRA-2011}.
 
\begin{figure}
\includegraphics[width=0.4\textwidth]{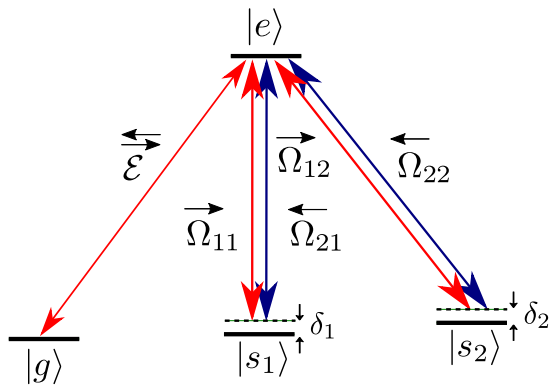}
\caption{(Color online) Tripod type linkage patterns for the creation of
two-component (spinor) slow-light by using two pairs of
counter-propagating control laser beams $\Omega_{j1}$ and $\Omega_{j2}$
(with $j=1,2$) driving atomic transitions from the unpopulated ground
states $|s_1\rangle$ and $|s_2\rangle$ to the excited state $|e\rangle$.
Two probe beams couple the populated atomic ground state $|g\rangle$ to
the excited state $|e\rangle$.}
\label{fig:fig2}
\end{figure}

In a recent letter \cite{Unanyan-PRL-2010} it has been demonstrated that
two-component slow light can be produced by means of a tripod scheme
which uses two standing wave control fields made of two pairs of
counter-propagating laser beams, as illustrated in Fig.~\ref{fig:fig2}.
Due to the formal similarity to two-component spinors we term the
two-component slow-light ''spinor'' slow light (SSL). We note however
that their transformation properties under Lorentz transformations are
not those of Dirac spinors. Employing two pairs of counterpropagating
beams involves two atomic coherences leading to the SSL.
By applying the secular approximation
\cite{Moiseev-PRA-2006,Zimmer-OC-2006}, the SSL has been shown to obey
an effective 1D Dirac equation \cite{Unanyan-PRL-2010}. This
approximation is however only justified in hot atomic gases
\cite{Bajcsy-Nature-2003,Fleischhauer-OC-1994}, because it neglects all
higher wave-vector components of the atomic coherence produced by the
counterpropagating beams driving the same transition. 

Here we study the propagation of two probe beams in an atomic ensemble
coherently driven by two pairs of counterpropagating control laser
fields in a double tripod-type linkage scheme shown in 
Fig.~\ref{fig:double-tripod}. In contrast to \cite{Unanyan-PRL-2010}
involving a single tripod scheme, no secular approximation is needed.
Thus the double tripod scheme can be used to produce SSL not only for
hot atomic gases but also for cold ones and in solids.  After
eliminating all atomic degrees of freedom and choosing proper amplitudes
and phases of the control lasers, the electric field strengths of the
SSL is described by an effective Dirac equation for a particle of finite
mass determined by the two photon detuning. The Dirac equation for
massive particles exhibits a finite energy gap given by the particles'
rest mass energy. Thus the atomic medium acts as a photonic crystal with
a controllable band gap. If the incoming probe light frequency lies
within the band gap, the light tunnels through the sample, with the
tunneling length being determined by the effective Compton length of the
SSL. On the other hand, for frequencies of the incoming probe light
outside the band gap, the transmission probability oscillates with
increasing length of the sample, so the system acts as a tunable filter
for certain frequencies. In both cases reflection takes place into the
complementary mode of the spinor probe field and thus is accompanied
by a change in frequency. Including the finite lifetime of the atomic
excited states leads to a loss term in the Dirac equation. We
investigate the influence of the decay on the transmission and
reflection of the SSL.

\section{Model}

\subsection{Double-tripod linkage pattern}

\begin{figure}
\includegraphics[height=0.3\textwidth]{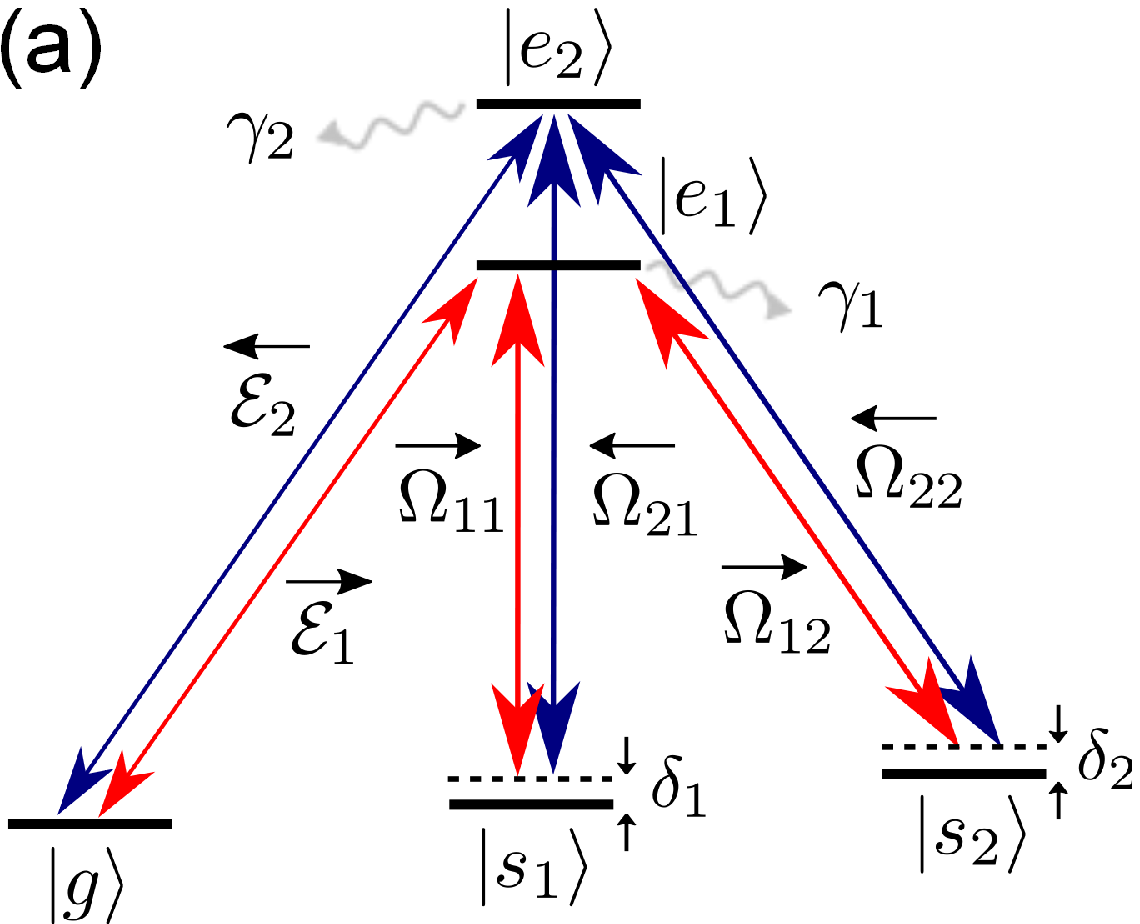}
\includegraphics[height=0.3\textwidth]{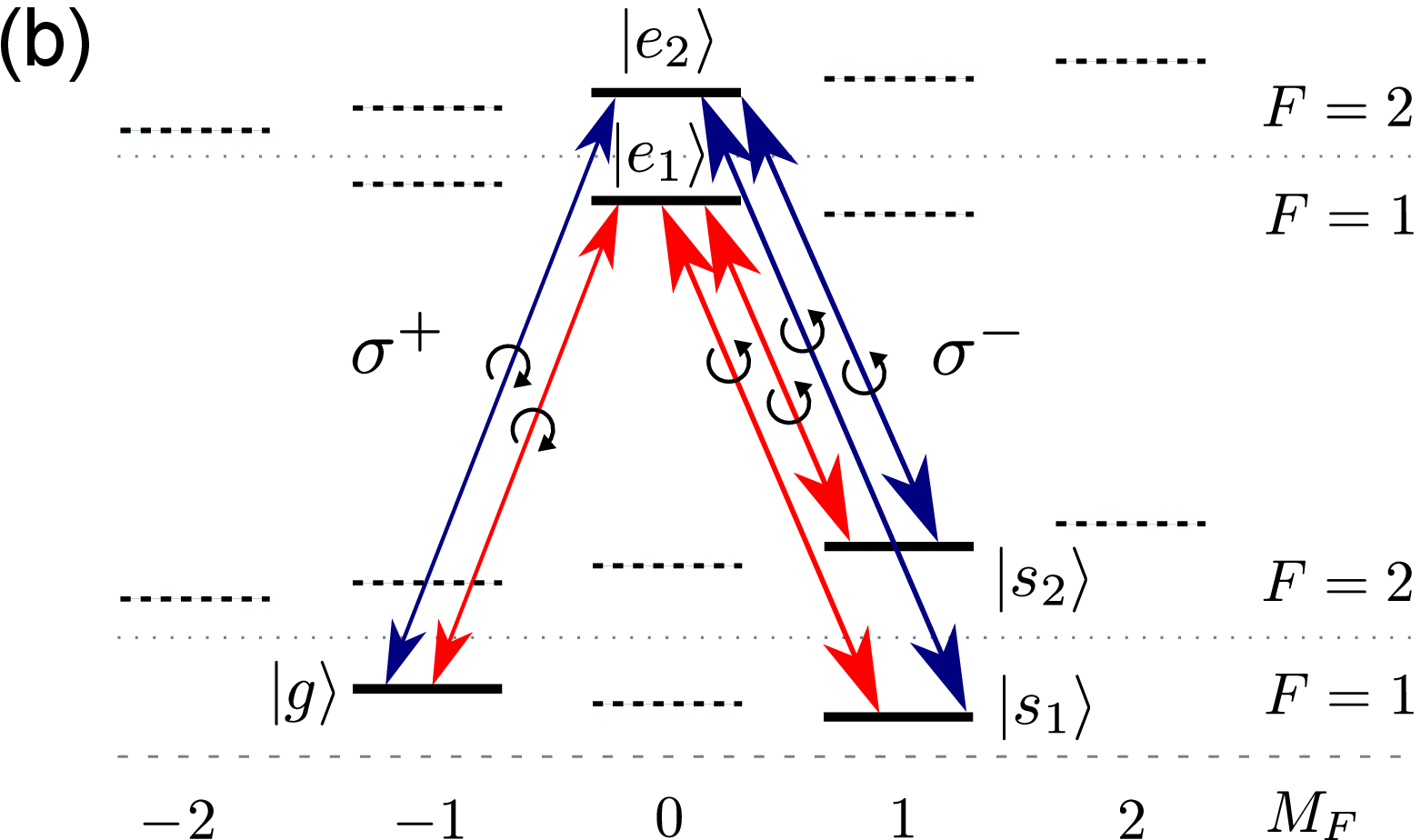}
\caption{(Color online) (a) Double tripod level structure for the
creation of spinor slow light. Two pairs of counter-propagating probe
fields characterized by the amplitudes $\mathcal{E}_1$ and
$\mathcal{E}_2$ couple resonantly the populated atomic ground state
$|g\rangle$ with two excited atomic states $|e_1\rangle$ and
$|e_2\rangle$. The propagation of the probe beams is controlled by two
pairs of counter-propagating control lasers beams (characterized by the
Rabi frequencies $\Omega_{j1}$ and $\Omega_{j2}$) driving the atomic
transitions from the excited state $|e_j\rangle$ to the unpopulated
ground states $|s_1\rangle$ and $|s_2\rangle$, with $j=1,2$. (b)
Possible experimental realization of the double tripod setup for atoms
like Rubidium \cite{Phillips-PRL-2001} or Sodium \cite{Liu-Nature-2001}.
The scheme involves transitions between the magnetic states of two
hyperfine levels with $F=1$ and $F=2$ for the ground and excited state
manifolds. Both probe beams are circular $\sigma^{+}$ polarized and all
four control beams are circular $\sigma^{-}$ polarized.}
\label{fig:double-tripod}
\end{figure}

We consider the propagation of two probe beams of light in a coherently
driven atomic ensemble exhibiting a double tripod level structure
depicted in Fig.~\ref{fig:double-tripod}a. The atoms are described by
three hyperfine ground levels $|g\rangle$, $|s_1\rangle$ and
$|s_2\rangle$ which are coupled to the electronic excited levels
$|e_1\rangle$ and $|e_2\rangle$ by probe (weaker) and control (stronger)
fields. Two probe beams $E_j$, $j=1,2$, with central frequencies
$\omega_1$ and $\omega_2$ are tuned to the atomic transitions
$|g\rangle\rightarrow|e_1\rangle$ and $|g\rangle\rightarrow|e_2\rangle$.
Four control laser beams couple two excited states $|e_j\rangle$ to
another two ground states $|s_q\rangle$, the coupling strength being
characterized by Rabi frequecies $\Omega_{jq}$, where $j,q=1,2$. The
control fields are strong enough to be treated as external parameters.
We assume four photon resonances between the probe beams and each pair
of the control lasers: $\omega_1-\omega_{1q}=\omega_2-\omega_{2q}$,
where $\omega_{jq}$ are the frequencies of the control fields. The
quantities $\Delta_j=\omega_{e_jg}-\omega_j$ and
$\delta_q=\omega_{s_qg}+\omega_{1q}-\omega_1=\omega_{s_qg}+\omega_{2q}-\omega_2$
define the one- and two-photon detuning from the one- and
two-photon resonances, respectively. Furthermore $\omega_{e_jg}$ and
$\omega_{s_qg}$ are the frequencies of the atomic transitions
$|g\rangle\rightarrow|e_j\rangle$ and $|g\rangle\rightarrow|s_q\rangle$.
In the following the control and probe beams are supposed to be close to
two-photon resonance. The simultaneous application of the probe and
control beams causes EIT in which the optical transitions from the
ground states interfere destructively thus preventing population of the
excited states $|e_1\rangle$ and $|e_2\rangle$. 

The double tripod scheme can be realized with atoms like Rubidium or
Sodium containing two hyperfine ground levels with $F=1$ and $F=2$, as
illustrated in Fig.~\ref{fig:double-tripod}b. These atoms have been
employed in the intial light-storage experiments based on a simpler
$\Lambda$ setup \cite{Liu-Nature-2001,Phillips-PRL-2001}. In the present
situation the states $|g\rangle$ and $|s_1\rangle$ correspond to the
magnetic sublevels with $M_F=-1$ and $M_F=1$ of the $F=1$ hyperfine
ground level, whereas the state $|s_2\rangle$ represents the hyperfine
ground state with $F=2$ and $M_F=1$. The two states $|e_1\rangle$ and
$|e_2\rangle$ correspond to the electronic excited states with $F=1$ and
$F=2$ characterized by $M_F=0$. To make a double tripod setup both probe
beams are to be circular $\sigma^{+}$ polarized and all four control
beams are to be circular $\sigma^{-}$ polarized. Note that such a scheme
can be implemented by adding three extra control laser beams as compared
to the experiment by Liu \textit{et al} \cite{Liu-Nature-2001}.

\subsection{Equation for the probe fields and atoms}

The electric field strength $E_j(\mathbf{r},t)$ of the $j$-th probe beam is
characterized by a slowly in time varying amplitude $\mathcal{E}_j$ normalized
to the number of photons:
\begin{equation}
E_j(\mathbf{r},t)=\sqrt{\frac{\hbar\omega}{2\varepsilon_0}}\mathcal{
E}_j(\mathbf{r},t)e^{-i\omega_jt}+\mathrm{c.c.}\,,\,\qquad j=1,2.
\label{eq:E-j}
\end{equation}
In the following we apply a semiclassical approach in which the dynamics of the
probe fields is described by classical Maxwell equations for the amplitudes
$\mathcal{E}_1$ and $\mathcal{E}_2$, whereas the atomic ensemble is described by
Schr\"odinger equations for the probability amplitudes (normalized to the atomic
density) $\Phi_g(\mathbf{r},t)$, $\Phi_{e_j}(\mathbf{r},t)$,
$\Phi_{s_q}(\mathbf{r},t)$ to find an atom at a position $\mathbf{r}$ in the
internal states $|g\rangle$, $|e_j\rangle$ and $|s_q\rangle$, respectively, with
$j,q=1,2$. It is convenient to write down the coupled light-matter equations of
motion in a matrix form. To this end we define  the two component spinors 
$\mathcal{E}=\left(\mathcal{E}_1,\mathcal{E}_2\right)^T$,
$\Phi_s=\left(\Phi_{s_1},\Phi_{s_2}\right)^T$ and
$\Phi_e=\left(\Phi_{e_1},\Phi_{e_2}\right)^T$. The following equation holds for
the slowly varying amplitudes of the probe fields:
\begin{equation}
\partial_t\mathcal{E}-\frac{i}{2}c\hat{k}^{-1}\nabla^2\mathcal{E}
-\frac{i}{2}c\hat{k}\mathcal{E}=ig\Phi_g^*\Phi_e
\label{eq:electric}
\end{equation}
where the r.h.s.\ of this equation is due to the atomic polarizability. Here
$\hat{k}=\text{diag}\left(k_j\right)$ is a diagonal $2\times2$ matrix with
elements $k_j=\omega_j/c$, $g=g_j=\mu_j(\omega_j/2\varepsilon_0\hbar)^{1/2}$
characterizes the atom-light coupling strength (assumed to be the same for both
probe fields) and $\mu_j$ is the dipole moment for the transition
$|g\rangle\rightarrow|e_j\rangle$. Neglecting effects due to atomic motion and
using the rotating wave approximation, the atomic probability amplitudes obey
the set of equations:
\begin{eqnarray}
i\hbar\partial_t\Phi_g &=&-\hbar g\mathcal{E}^{\dag}\Phi_e \label{eq:phig} \\
i\hbar\partial_t\Phi_e &=&\hbar\hat{\Delta}\Phi_e-\hbar\hat{\Omega}\Phi_s
-\hbar g\Phi_g\mathcal{E} \label{eq:e} \\
i\hbar\partial_t\Phi_s &=&\hbar\hat{\delta}\Phi_s
-\hbar\hat{\Omega}^{\dag}\Phi_e \label{eq:s}
\end{eqnarray}
where $\hat{\Omega}$ is a $2\times 2$ matrix of Rabi frequencies $\Omega_{ij}$,
and the dagger refers to a Hermitian conjugated matrix. On the other hand,
$\hat{\Delta}$ and $\hat{\delta}$ are the following diagonal $2\times 2$
matrices
\begin{equation}
\hat{\Delta}=\left(
\begin{array}{cc}
\Delta_1-i\gamma_1 & 0\\ 0 &\Delta_2-i\gamma_2
\end{array}\right)\,,\quad\hat{\delta}=\left(
\begin{array}{cc}
\delta_1 & 0\\ 0 &\delta_2
\end{array}\right)
\label{eq:Delta--delta}
\end{equation}
where $\Delta_j$ and $\delta_j$ are the detunings from the one- and two-photon
resonances, respectively, and $\gamma_j$ is the decay rate of the $j$-th excited
electronic level. Note that the appearance of non-zero decay rates should
generally be accompanied by introducing noise operators in the equations of
motion \cite{Scully-Book-1997}. Yet in the present situation one can disregard
the latter noise, since we are working in the linear regime with respect to the
probe field leading to a negligible population of the excited state. Assuming
that the inverse matrix $\hat{\Omega}^{-1}$ exists and using Eq.~(\ref{eq:s}) one
can relate $\Phi_e$ to $\Phi_s$ and obtain
\begin{equation}
\Phi_e=(\hat{\Omega}^{\dag})^{-1}\left[-i\partial_t+\hat{\delta}\right]\Phi_s\,,
\label{eq:e-s}
\end{equation}
On the other hand, Eq.~(\ref{eq:e}) relates the atomic coherence $\Phi_s$ to
the probe field $\mathcal{E}$ as:
\begin{equation}
\Phi_s=-g\Phi_g\hat{\Omega}^{-1}\mathcal{E}+\hat{\Omega}^{-1}\left(
\hat{\Delta}-i\partial_t\right)\Phi_e\,.
\label{eq:s-general}
\end{equation}
The last equation will serve as a starting point for the adiabatic approach.
It should be noted that one can also treat the case when
$\hat\Omega$ is a singular matrix by computing, e.g., the Moore-Penrose
pseudo-inverse \cite{Laub-2004}. This case however is of no interest here,
since it results in an effective double-$\Lambda$ system.

\section{Equations for spinor slow light}

\subsection{Adiabatic elimination of the excited states}
The zero-order adiabatic approximation is obtained by neglecting the populations
of the excited states in Eq.~(\ref{eq:s-general}), giving
\begin{equation}
\Phi_s=-g\Phi_g\hat{\Omega}^{-1}\mathcal{E}\,.
\label{eq:s-electric}
\end{equation}
The higher order corrections will be considered later in
Sec.~\ref{sec:Dirac-equation-with} when treating the effects of finite excited
state lifetimes. Initially all atoms are assumed to be in the ground level
$|g\rangle$. As the Rabi frequencies of the probe fields are much smaller than
those of the control fields, one can neglect the depletion of the ground level
$|g\rangle$, the population of the latter determining the atomic density
$n=|\Phi_g|^2$. Using Eqs.~(\ref{eq:e-s}) and (\ref{eq:s-electric}) and taking
$\Phi_g=\sqrt{n}$ one can eliminate the atomic spin coherence $\Phi_s$ and
express the excited-state amplitudes via the amplitudes of the probe fields:
\begin{equation}
\Phi_e=g(\hat{\Omega}^{\dag})^{-1}\left[i\partial_t-\hat{\delta}\right]
(n^{1/2}\hat{\Omega}^{-1}\mathcal{E}\,).
\label{eq:e-E}
\end{equation}
Equations (\ref{eq:electric}) and (\ref{eq:e-E}) provide a closed set of
equations for the electric field amplitudes $\mathcal{E}_1$ and $\mathcal{E}_2$
of the SSL.

In the following the pairs of the control beams $\Omega_{1j}$ and $\Omega_{2j}$
are taken to counter-propagate along the $z$ axis:
$\Omega_{1j}=\tilde{\Omega}_{1j}e^{ik_{1j}z}$,
$\Omega_{2j}=\tilde{\Omega}_{2j}e^{-ik_{2j}z}$, where $k_{ij}$ are the wave
numbers of the control beams characterized by the amplitudes
$\tilde{\Omega}_{ij}$, with $i,j=1,2$. The probe fields also counter-propagate
along the $z$ axis:
$\mathcal{E}_1(\mathbf{r},t)=\tilde{\mathcal{E}}_1(\mathbf{r},t)e^{ik_1z}$ ,
$\mathcal{E}_2(\mathbf{r},t)=\tilde{\mathcal{E}}_2(\mathbf{r},t)e^{-ik_2z}$,
with $k_j=\omega_j/c$ being the central wave-vector of the $j$-th probe beam.
For paraxial beams $\tilde{\mathcal{E}}_1(\mathbf{r},t)$ and
$\tilde{\mathcal{E}}_2(\mathbf{r},t)$ represent the slowly varying amplitudes
which depend weakly on the propagation direction $z$. Furthermore we assume
that $k_1\approx k_{11}\approx k_{12}$ and $k_2\approx k_{21}\approx k_{22}$.
We take the amplitudes of the control beams $\tilde{\Omega}_{ij}$ to be
time-independent, neglect their position-dependence and assume the atomic
density to be homogeneous throughout the sample. The slowly varying
two-component amplitude
$\tilde{\mathcal{E}}=(\tilde{\mathcal{E}}_1\,,\tilde{\mathcal{E}}_2)^T$ obeys
the following paraxial equation:
\begin{equation}
\sigma_z(c^{-1}+\tilde{v}^{-1})\partial_t\tilde{\mathcal{E}}
+\partial_z\tilde{\mathcal{E}}=
i\sigma_z(2\hat{k})^{-1}\nabla_{\bot}^2\tilde{\mathcal{E}}
-i\sigma_z\tilde{v}^{-1}\tilde{D}\tilde{\mathcal{E}},
\label{eq:result3}
\end{equation}
where
\begin{equation}
\tilde{D}=\tilde{\Omega}\hat{\delta}\tilde{\Omega}^{-1}\,,
\end{equation}
$\tilde{\Omega}$ is a $2\times 2$ matrix with matrix elements
$\tilde{\Omega}_{ij}$, $\sigma_z$ is a Pauli matrix and $\sigma_z\tilde{v}^{-1}$
represents the inverse group velocity matrix of slow light with
\begin{equation}
\tilde{v}^{-1}=\frac{g^2n}{c}(\tilde{\Omega}^{\dag})^{-1}\tilde{\Omega}^{-1}\,.
\label{eq:sigma-z-v}
\end{equation}
From now on the Rabi frequencies of the control beams
$\tilde{\Omega}_{ij}=|\Omega_{ij}|e^{iS_{ij}}$ are considered to have the same
amplitudes: $|\tilde{\Omega}_{ij}|=\Omega /\sqrt{2}$ and tunable phases
$S_{ij}\,$. The latter $S_{ij}$ can be made to be $S_{11}=S_{22}=0$ and
$S_{12}=S_{21}=S$ by properly choosing the phases of the atomic and radiation
fields. Thus one has 
\begin{equation}
\tilde{\Omega}=\frac{\Omega}{\sqrt{2}}\left(I+e^{iS}\sigma_x\right)\,.
\label{eq:Omega-tilde}
\end{equation}
Equations (\ref{eq:sigma-z-v})--(\ref{eq:Omega-tilde}) yield
\begin{equation}
\sigma_z\tilde{v}^{-1}=\frac{1}{v_0\sin^2S}
\left(\sigma_z-i\cos S\sigma_y\right)
\label{eq:sigma-v-z-1}
\end{equation}
where
\begin{equation}
v_0=\frac{c\Omega^2}{g^2n}
\label{eq:v-0}
\end{equation}
is the group velocity of slow light. Furthermore by taking the detunings to have
the opposite signs $\delta_2=-\delta_1\equiv\delta$, the matrix $\tilde{D}$
simplifies to
\begin{equation}
\tilde{D}=\frac{\delta}{\sin S}
\left(i\cos S\sigma_z+\sigma_y\right).
\label{eq:D-tilde-explicit}
\end{equation}
It should be noted that by changing relative phase $S$ one can
considerably alter the time evolution of the SSL. For zero two photon
detuning, i.e.\ $\delta=0$, the case of $S=\pi/2$ corresponds to two
independent tripod schemes, whereas in the limit $S\rightarrow 0$ one
recovers the double-$\Lambda$ scheme, as $\hat\Omega$ becomes singular.

\subsection{Paraxial Dirac equation}

\begin{figure}
\includegraphics[width=0.6\textwidth]{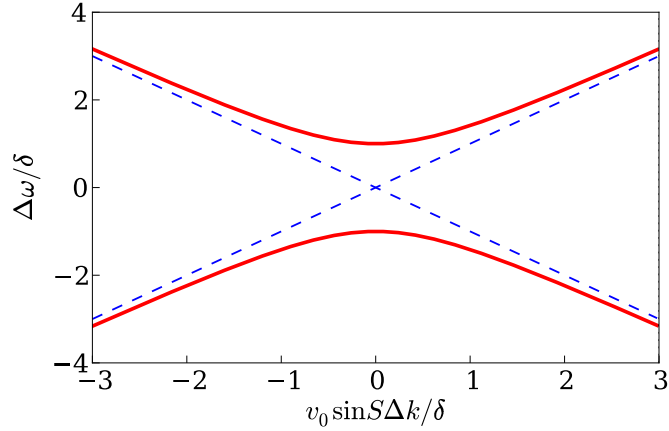}
\caption{(Color online) Dirac dispersion of slow light for non-zero
two-photon detuning $\delta\neq 0$  (solid red line) together with the
asymptotic behavior at large $\Delta k$ (dashed blue line).}
\label{fig:dispersion}
\end{figure}

Neglecting diffraction effects and using
Eqs.~(\ref{eq:sigma-v-z-1})--(\ref{eq:D-tilde-explicit}), the equation of motion
(\ref{eq:result3}) takes the form
\begin{equation}
\bigg[\bigg(\frac{1}{c}+\frac{1}{v_0}\frac{1}{\sin^2S}\bigg)\sigma_z-
i\frac{1}{v_0}\frac{\cos S}{\sin^2S}\sigma_y\bigg]
\frac{\partial}{\partial t}\tilde{\mathcal{E}}+\frac{\partial}{\partial z}\tilde{\mathcal{E}}=
-\frac{\delta}{v_0\sin S}\sigma_x\tilde{\mathcal{E}}\,.
\label{eq:pol-2}
\end{equation}
In the regime of slow light one has $v_0/c\ll 1$. In such a case, taking
$S=\pi/2$, the above equation reduces to the following Dirac equation
for a massive particle:
\begin{equation}
(i\partial_t+iv_0\sigma_z\partial_z-\delta\sigma_y)\tilde{\mathcal{E}}=0\,.
\label{eq:pol-eq-pi/2}
\end{equation}
Assuming a monochromatic probe field
$\tilde{\mathcal{E}}\sim e^{-i\Delta\omega t}$,
it is convenient to rewrite Eq.~(\ref{eq:pol-2}) in terms of a complex
vector $\mathbf{K}=\left(iK_x,iK_y,K_z\right)$ and the vector of Pauli matrices
$\boldsymbol{\sigma}=\left(\sigma_x,\sigma_y,\sigma_z\right)$
\begin{equation}
\partial_z\tilde{\mathcal{E}}=i\mathbf{K}\cdot
\boldsymbol{\sigma}\tilde{\mathcal{E}},
\label{eq:pol-short}
\end{equation}
with
\begin{equation}
K_x = \frac{\delta}{v_0\sin S},\qquad
K_y = -\frac{\Delta\omega\cos S}{v_0\sin^2S},\qquad
K_z = \frac{\Delta\omega}{c}+\frac{\Delta\omega}{v_0\sin^2S}.
\label{eq:Kxyz}
\end{equation}
Equation (\ref{eq:pol-short}) has plane wave solutions
$\tilde{\mathcal{E}}=\chi e^{i\Delta kz}$, where the column $\chi$ obeys the
eigenvalue equation $\Delta k\chi =\mathbf{K}\cdot \boldsymbol{\sigma}\chi$.
Eigenvalues of the matrix $\mathbf{K}\cdot\boldsymbol{\sigma}$ are $\pm K$,
where
\begin{equation}
K=\sqrt{K_z^2-K_x^2-K_y^2}
\end{equation}
is the length of the complex vector $\mathbf{K}$. Thus the dispersion is
given by $\Delta k^2=K^2$. For the slow light, $v_0/c\ll 1$, one obtains
\begin{equation}
\Delta\omega^{\pm}=\pm\sqrt{\delta^2+\Delta k^2v_0^2\sin^2S}\,.
\label{eq:dispersion-slow-light}
\end{equation}
Equation (\ref{eq:dispersion-slow-light}) is analogous to the dispersion
of a relativistic particle with an effective mass
$m=\hbar\delta/(v_0\sin S)^2$. The latter is determined by the
two-photon detuning $\delta$ and the relative phase of the control beams
$S$. The effective speed of light is given by the velocity $v_0|\sin
S|$. At small $S$ we have quadratic dispersion characteristic to
stationary light \cite{Zimmer-OC-2006,Fleischhauer-PRL-2008}. As
illustrated in Fig.~\ref{fig:dispersion}, the two dispersion branches
with positive and negative effective mass are separated by a gap
$\delta$. Thus the atomic medium acts as a photonic crystal with a
controllable band gap. For $|\Delta\omega |<\delta$ the eigenfunctions
become evanescent and are characterized by an imaginary wave vector
$\Delta k=i\Delta q$. Consequently there are no propagating waves in
this range, resulting in the formation of a band-gap.

\section{Reflection and transmission of the probe beam}

\begin{figure}
\includegraphics[width=0.4\textwidth]{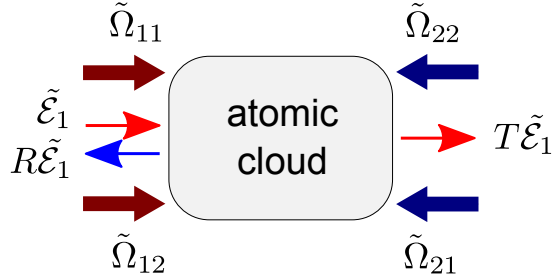}
\caption{(Color online) Transmission and reflection of the incident probe
field $\tilde{\mathcal{E}}_1=\tilde{\mathcal{E}}_1(0,t)$. The
transmitted field is given by $\tilde{\mathcal{E}}_1(L,t)=T\tilde{\mathcal{E}}_1(0,t)$, whereas
the reflected part is determined by
$\tilde{\mathcal{E}}_2(L,t)=R\tilde{\mathcal{E}}_1(0,t)$, where $T$,
$R$ denote the transmission and reflection coefficients, respectively.}
\label{fig:reflection-setup}
\end{figure}

Let us analyze the transmission of a probe beam through the atomic
cloud, as well as the accompanying reflection. The atomic gas is
considered to be uniform along the propagation direction $z$ from the entry
point of the probe beam at $z=0$ to its exit at $z=L$. The incoming probe
field contains the first component $\tilde{\mathcal{E}}_1(z,t)$ and is
monochromatic
$\tilde{\mathcal{E}}_1(0,t)=\tilde{\mathcal{E}}_0e^{-i\Delta\omega t}$
with the frequency detuned from the central frequency $\omega_1$ by the
amount $\Delta\omega$, where $\tilde{\mathcal{E}}_0$ is the amplitude of
the incoming field. As illustrated in Fig.~\ref{fig:reflection-setup},
the probe field is transmitted through the atomic cloud with the
amplitude $T$ and is reflected to the second component with the
amplitude $R$,
i.e.\ $\tilde{\mathcal{E}}_1(L,t)=T\tilde{\mathcal{E}}_1(0,t)$ and
$\tilde{\mathcal{E}}_2(0,t)=R\tilde{\mathcal{E}}_1(0,t)$. This leads to
the following boundary conditions for the two-component probe field
\begin{eqnarray}
\tilde{\mathcal{E}}(0,t) & = &\tilde{\mathcal{E}}_0\left(
\begin{array}{c}
1\\ R
\end{array}\right)e^{-i\Delta\omega t},\\
\tilde{\mathcal{E}}(L,t) & = &\tilde{\mathcal{E}}_0\left(
\begin{array}{c} T\\
0
\end{array}\right)e^{-i\Delta\omega t}.
\label{eq:Boundary-conditions}
\end{eqnarray}
The spatial development of monochromatic probe fields is described by
Eq.~(\ref{eq:pol-short}) with the formal solution
$\tilde{\mathcal{E}}(z,t)=e^{i\mathbf{K}\cdot\boldsymbol{\sigma}z}\tilde{
\mathcal{E}}(0,t)$. Thus one can relate the two-component probe field at the
entrance and exit points by
\begin{equation}
\tilde{\mathcal{E}}(L,t)=\left[\cos(KL)+i\frac{\mathbf{K}}{K}\cdot
\boldsymbol{\sigma}\sin(KL)\right]\tilde{\mathcal{E}}(0,t)\,.
\label{eq:sol}
\end{equation}
Combining Eqs.~(\ref{eq:Boundary-conditions}) and (\ref{eq:sol}) one finds the
reflection and transmission coefficients
\begin{eqnarray}
R &=&\frac{\left(K_x+iK_y\right)\sin(KL)}{K\cos(KL)-iK_z\sin(KL)}, \label{eq:R}\\
T &=&\frac{K}{K\cos(KL)-iK_z\sin(KL)}\,,
\label{eq:T}
\end{eqnarray}
where the general expressions for $K_{x,y,z}$ are given in
Eq.~(\ref{eq:Kxyz}). In what follows we are interested in the regime of slow
propagation of the probe light within the atomic cloud ($v_0\ll c$). In this
case  $K_z$ simplifies to $K_z\approx\Delta\omega/v_0\sin^2S$ and thus
\begin{equation}
K=\frac{1}{v_0|\sin S|}\sqrt{\Delta\omega^2-\delta^2}\,.
\label{eq:K-slow-light}
\end{equation}
From Eq.~(\ref{eq:Boundary-conditions}) it becomes clear that the reflection
takes place into the complementary mode of the probe field and can be
accompanied by a change in frequency, as the center frequencies $\omega_j$ of
the probe fields do not have to be equal.

\subsection{Oscillations of transition and reflection amplitudes}

\begin{figure}
\includegraphics[width=0.3\textwidth]{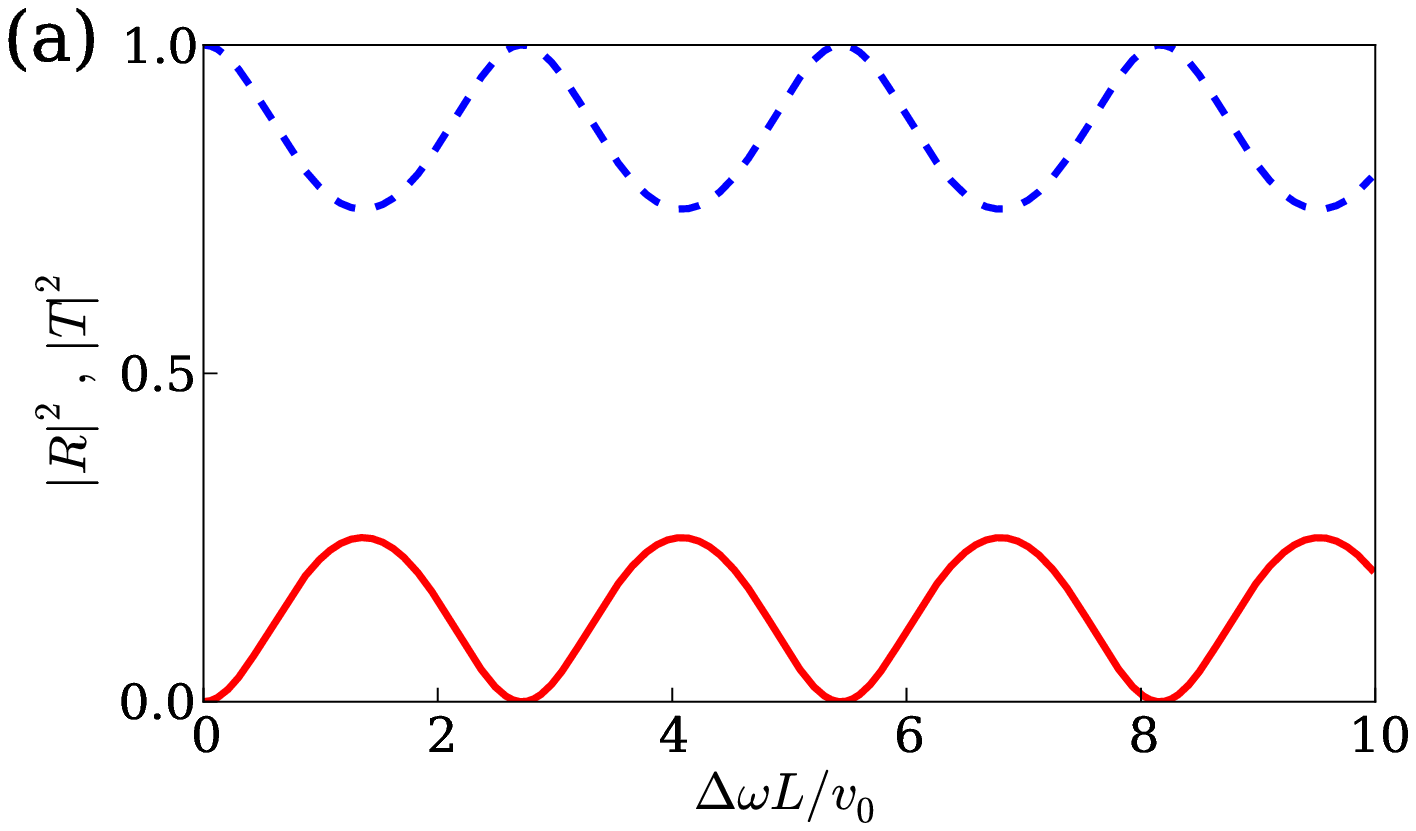}
\includegraphics[width=0.3\textwidth]{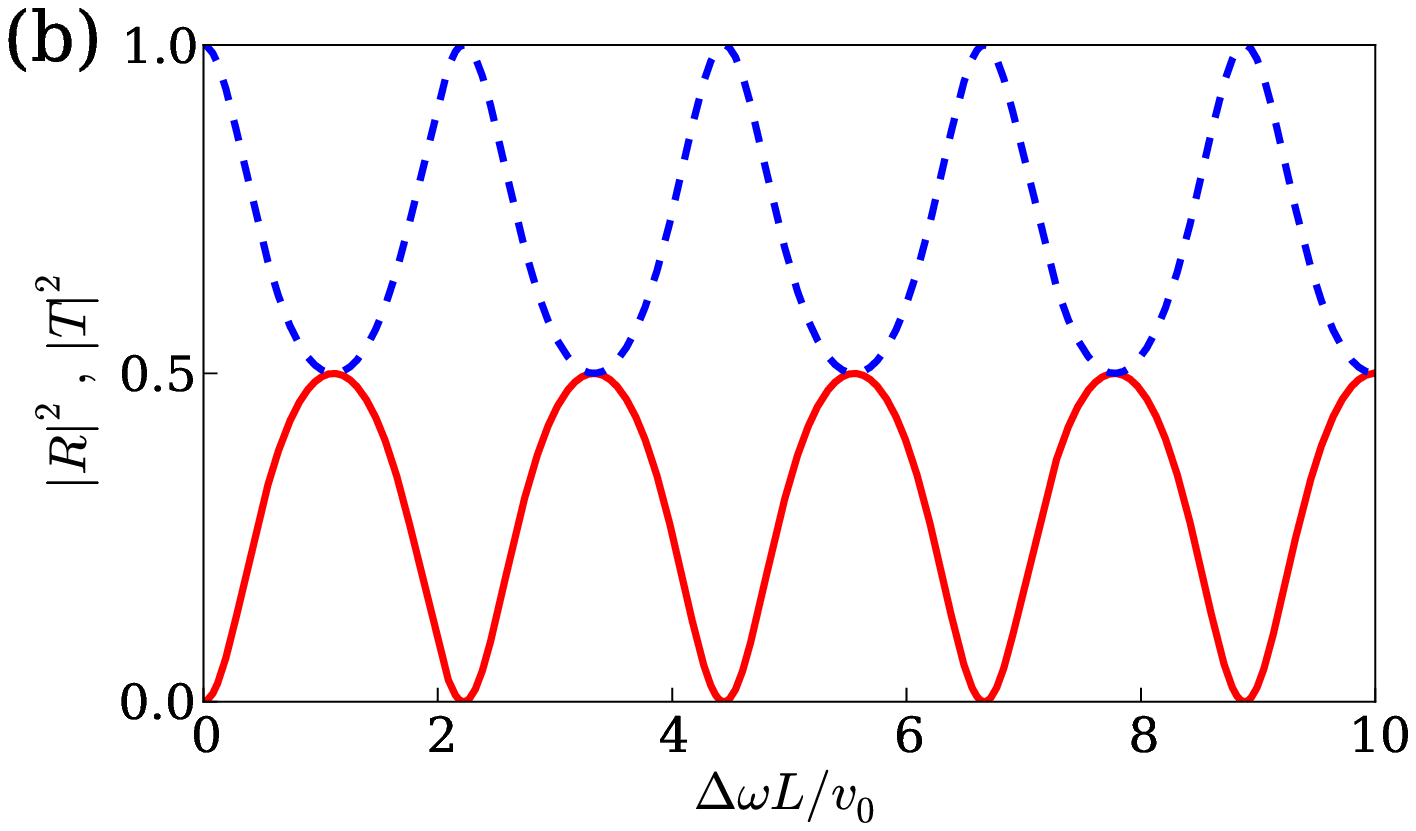}
\includegraphics[width=0.3\textwidth]{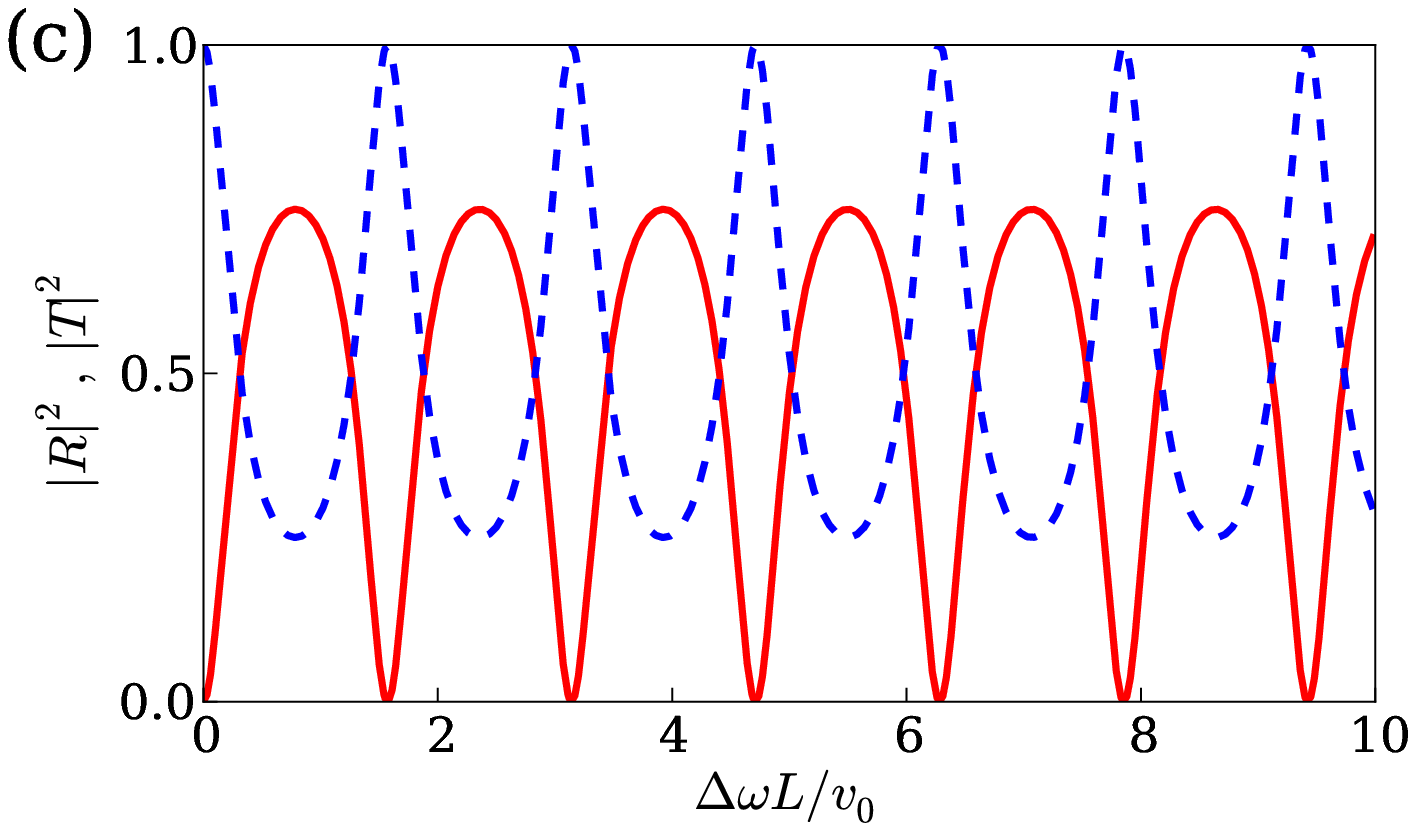}
\caption{(Color online) Dependence of the reflection and transmission
probabilities $|R|^2$ and  $|T|^2$   (shown in solid red and dashed blue
lines respectively) on the dimensionless sample length $\Delta\omega
L/v_0$ for zero two-photon detuning ($\delta=0$) in the case
where $S=\pi/3$ (a), $S=\pi/4$ (b) and $S=\pi/6$ (c).}
\label{fig:oscillations}
\end{figure}

For probe light frequencies outside the band gap
$\left(\Delta\omega\right)^2>\delta^2$, the transmission and reflection
amplitudes oscillate with increasing system length. Such a behavior is
characteristic to light passing through resonant cavities. Thus the system acts
as a frequency filter without mirrors. For zero two-photon detuning
($\delta=0$), the transmission and reflection amplitudes (\ref{eq:R}) and (\ref{eq:T}) simplify to
\begin{equation}
R=-\frac{i\cos S\sin(KL)}{|\sin S|\cos(KL)-i\sin(KL)},
\label{eq:R-zero-delta}
\end{equation}
and
\begin{equation}
T=\frac{|\sin S|}{|\sin S|\cos(KL)-i\sin(KL)},
\label{eq:T-zero-delta}
\end{equation}
with $K=\Delta\omega/v_0|\sin S|$. Fig.~\ref{fig:oscillations}
illustrates the oscillatory behavior of the transmission and reflection
probabilities $|T|^2$, $|R|^2$ on the sample length $L$ for zero
two-photon detuning ($\delta =0$) and non-zero detuning
$\Delta\omega\neq 0$ of the incident probe field. The complete transfer
of the probe field through the sample occurs at
$\Delta\omega L=\pi j v_0|\sin S|$, with $j$ being an integer.
The frequency difference between two such resonance maxima
$\pi v_0|\sin S|/L$ is inversely proportional to the sample length $L$.
For instance, if we take the group velocity of slow light
$v_0=17\,\mathrm{m/s}$ and the length of the atomic cloud
$L=300\,\mu\mathrm{m}$ as in the experiments \cite{Hau-Nature-1999,Liu-Nature-2001}
and choose $S=\pi /4$, the period of the oscillations $\Delta\omega$
is around $10^5\,\mathrm{Hz}$. Note that the minima of the transfer
amplitude $|T|=|\sin S|$ correspond to
$\Delta\omega L=\pi(j+1/2)v_0|\sin S|$. Thus the reflection coefficient
$R$ oscillates from $0$ to the maximum value $|\cos S|$. The transmission
and reflection coefficients $T$ and $R$ are seen to be sensitive to the
relative phase of the laser beams $S$. In the limit $S\rightarrow 0$ the
transfer probability is approaching zero ($T\rightarrow 0$) which is
accompanied by a complete reflection to the second field,
i.e.\ $|R|\rightarrow 1$. This corresponds to the creation of a photonic
band-gap \cite{Andre-2002,Fleischhauer-PRL-2008} in the resulting double
$\Lambda$ scheme. For $S=\pm\pi /2$ the reflection is zero ($R=0$) and
there is a complete transfer of the original field through the sample
($|T|=1$). In that case the double tripod reduces to two independent
tripod schemes. Introducing a small two-photon detuning $\delta\neq 0$
mixes the two counter-propagating probe field components, leading to a
non-zero reflection ($R\neq 0$) even for $S=\pm\pi/2$.

\subsection{Tunneling of slow light}

\begin{figure}
\includegraphics[width=0.6\textwidth]{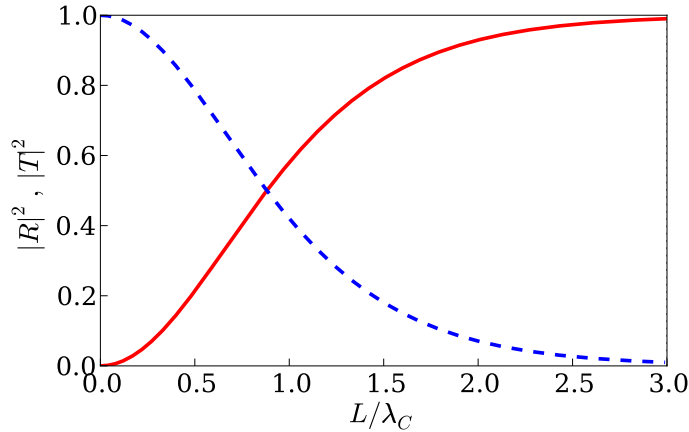}
\caption{(Color online) Dependence of the reflection and transmission
probabilities $|R|^2$ and $|T|^2$ (shown in solid red and dashed blue
lines respectively) on the dimensionless two-photon detuning $L/ \lambda_{C}=\delta
L/(v_0|\sin S|)$ for $\Delta\omega=0$.}
\label{fig:transmission}
\end{figure}

In the case where the probe light frequency lies within the band-gap
$\left(\Delta\omega\right)^2<\delta^2$), the wave-number $K=i|K|$
becomes imaginary. In such a situation, Eq.~(\ref{eq:T}) describes the
decay of the transmission amplitude with distance. In particular, for
$\Delta\omega =0$ the reflection and transmission amplitudes
(\ref{eq:R}) and (\ref{eq:T}) simplify to
\begin{equation}
R=\tanh(K_xL)\,,\qquad T=\frac{1}{\cosh(K_xL)}
\label{eq:T-T-0}
\end{equation}
with $K_x=\delta/v_0\sin S$. The dependence of the reflection and transmission
probability on the product of detuning and sample length is presented in
Fig.~\ref{fig:transmission}. The light tunnels through the sample, the tunneling
length being determined by the effective Compton length
\begin{equation}
\lambda_{\mathrm{C}}=v_0|\sin S|/\delta\,.
\label{eq:lambda-Compton}
\end{equation}
In fact, a relativistic particle is known to be characterized by a
Compton wavelength $\lambda_{\mathrm{C}}=\hbar /mc$. In the present
situation the Compton length reads $\lambda_{\mathrm{C}}=\hbar
/(mv_0|\sin S|)$, where $m=\hbar\delta /(v_0\sin S)^2$ is the effective
mass of the SSL. Using Eq.~(\ref{eq:T-T-0}) one can see that the
transmission of the incident wave is efficient as long as the length of
the gas cloud $L$ is much smaller than the Compton wavelength:
$L\ll\lambda_{\mathrm{C}}$. For larger values of $L$ the transmission
probability falls off exponentially. This behavior is related to the
fact that it is impossible to localize a particle with an uncertainty
smaller than the Compton wavelength \cite{Unanyan-2009,Toyama-2010}.
Since the Compton length can be tuned by changing $\delta$ it is
possible to experimentally study the tunneling regime $L\ll \lambda_C$.

If we take the length of atomic cloud to be $L=0.3\,\mathrm{mm}$ and the
group velocity of light is $v_0=17\,\mathrm{m/s}$
\cite{Hau-Nature-1999}, the Compton length becomes of the order of the
length of the atomic cloud, when the detuning is equal to
$\delta_1=v_0/L\approx6\times10^4\,\mathrm{Hz}$. This is well within
the EIT transparency window which is of the order of a $1\,\mathrm{MHz}$
in the experiment \cite{Hau-Nature-1999} and absorption losses due to a
finite two-photon detuning can be neglected. In contrast to ordinary
absorption, the decrease in the transmission of light through the sample
is now accompanied by an increase in the reflection into the
complementary mode satisfying the unitarity condition $|R|^2+|T|^2=1$.

\section{Influence of losses \label{sec:Dirac-equation-with}}

Let us now analyze the losses due to the finite lifetime of the excited states.
For this we take into account the next order of iteration in
Eq.~(\ref{eq:s-general}) and include the decay rates $\gamma_1$ and $\gamma_2$
in the matrix $\hat{\Delta}$. Assuming the decay
rates to be the same for both excited states ($\gamma_1=\gamma_2\equiv\gamma$)
and putting to zero  the one-photon detunings $\Delta_1=\Delta_2=0$, 
the r.h.s.\ of the general equation of motion
(\ref{eq:result3}) acquires an extra term
\begin{equation}
i\sigma_z\frac{\gamma c}{g^2n}\left(\tilde{v}^{-1}\tilde{D}\right)^2
\mathcal{E}.
\end{equation}
In the case of slow light ($v_0/c\ll 1$) and neglecting
diffraction effects one has for $S=\pi/2$
\begin{equation}
\partial_t\tilde{\mathcal{E}}
+v_0\sigma_z\partial_z\tilde{\mathcal{E}}
+i\delta\sigma_y\tilde{\mathcal{E}}+\gamma_\mathrm{eff}\tilde{\mathcal{E}}=0\,,
\label{eq:Eq-no-difraction}
\end{equation}
where $\gamma_\mathrm{eff}=\gamma\delta^2/\Omega^2$ is the effective decay rate of
the probe light fields.

Eq.~(\ref{eq:Eq-no-difraction}) represents a one-dimensional Dirac equation with
losses which extends the previous equation (\ref{eq:pol-eq-pi/2}). As a result,
one needs to replace $\Delta\omega$ by
$\Delta\omega^{\prime}=\Delta\omega-i\gamma_\text{eff}$ in the corresponding
reflection and transmission coefficients. For zero probe field detuning
($\Delta\omega =0$) and $v_0\ll c$, the transmission and reflection coefficient take the form
\begin{eqnarray}
T &=&\frac{\delta_\mathrm{eff}}{\delta_\text{eff}\cosh\left(\frac{L}{v_0}\delta_\mathrm{eff}\right)
+\gamma_\mathrm{eff}\sinh\left(\frac{L}{v_0}\delta_\mathrm{eff}\right)},\\
R &=&\frac{\delta\sinh\left(\frac{L}{v_0}\delta_\mathrm{eff}\right)}{\delta_\mathrm{eff}
\cosh\left(\frac{L}{v_0}\delta_\mathrm{eff}\right)
+\gamma_\mathrm{eff}\sinh\left(\frac{L}{v_0}\delta_\mathrm{eff}\right)},
\end{eqnarray}
where we defined
\begin{equation}
\delta_\mathrm{eff}=\sqrt{\delta^2+\gamma_\mathrm{eff}^2}\,.
\end{equation}
For a large sample size
$L\gg v_0/\delta_\mathrm{eff}$ these equations simplify to
\begin{eqnarray}
T &\approx &\frac{2\delta_\mathrm{eff}}{\delta_\mathrm{eff}+\gamma_\mathrm{eff}}
\exp\left(-\frac{L}{v_0}\delta_\mathrm{eff}\right)\,, \label{eq:Trans} \\
R &\approx &\frac{\delta}{\delta_\mathrm{eff}+\gamma_\mathrm{eff}}.
\label{eq:Reflect}
\end{eqnarray}
The transmission coefficient $T$ decays exponentially with the system
length $L$, while the reflection coefficient stays non-zero even for
infinitely long samples. For sufficiently small detuning the EIT
condition \cite{Fleischhauer-RMP-2005} is fulfilled
$\gamma\delta/\Omega^2\ll 1$. Thus one arrives at an almost perfect reflection
$R\approx 1-\delta\gamma/\Omega^2\approx 1$. In the opposite case 
$\gamma\delta /\Omega^2\gg1$, the EIT condition is violated and the
probe fields experience strong losses leading to vanishing reflectivity.
This is related to the fact that for $\gamma_\text{eff}\neq 0$ the
unitarity condition is violated  $|R|^2+|T|^2<1$ leading to the reduced
reflectivity.

\section{Conclusions}

We studied two component (spinor) slow light in an ensemble of
atoms coherently driven by two pairs of counter-propagating control
laser fields in a double tripod-type linkage scheme. The SSL obeys an
effective Dirac equation for a massive particle. By changing the
two-photon detuning the atomic medium can act as a photonic crystal with
a controllable band-gap. This gap is equivalent to the rest mass energy
splitting in the Dirac dispersion. We investigated the dependence of
tunneling and transmission rates of the incoming probe fields on its
frequency. For frequencies within the band-gap the probe light tunnels
through the sample with the tunneling length given by the effective
Compton wave-length of the SSL. In the case of a sample length exceeding
the Compton wave length of the SSL ($L\gg\lambda_C$), the formation of
the band-gap leads to the perfect reflection. In the opposite limit of a
short sample length ($L\ll\lambda_C$) the transmission probability is
close to unity, as the SSL can not be localized below the Compton
wave-length. For frequencies of the probe light outside the band-gap,
the reflection and transmission coefficients exhibit an oscillatory
dependence on the two-photon detuning and the sample length. This can be
interpreted as a mirrorless frequency filter.

We discussed the effect of finite excited state lifetimes on
transmission and reflection. For sufficiently small loss rates the
reflection and transmission coefficients fulfill the unitarity condition,
and the reflection takes place into the complementary mode of the SSL.
Increasing the loss rates leads to non-adiabatic losses and the
unitarity condition no longer holds. Finally, we proposed a possible
experimental realization of two-component slow light using double
tripod scheme with alkali atoms like Rubidium or Sodium.

\begin{acknowledgments}
This work has been supported by the Research Council of Lithuania
(grants No.\ MOS-13/2011 and VP1-3.1-\v{S}MM-01-V-01-001),
the EU project NAMEQUAM, and the DFG projects UN 280/1 and GRK 792.
\end{acknowledgments}

\end{document}